\documentclass[reprint,
 amsmath,amssymb,
 aps,
pra,
floatfix,
]{revtex4-1}
\usepackage{graphicx}
\usepackage{dcolumn}
\usepackage{bm}
\usepackage{siunitx}
\begin{document}
\preprint{APS/123-QED}

\title{Characterization of laser-induced ionization dynamics in solid dielectrics}

\author{Peter J{\"u}rgens$^{1}$}
\email{juergens@mbi-berlin.de}
\author{Benjamin Liewehr$^{2}$}
\author{Bj{\"o}rn Kruse$^{2}$}
\author{Christian Peltz$^{2}$}
\author{Tobias Witting$^{1}$}
\author{Anton Husakou$^{1}$}
\author{Arnaud Rouze{\'e}$^{1}$}
\author{Mikhail Ivanov$^{1}$}
\author{Thomas Fennel$^{2}$}
\author{Marc J. J. Vrakking$^{1}$}
\author{Alexandre Mermillod-Blondin$^{1}$}

\affiliation{$^{1}$Max-Born-Institute for Nonlinear Optics and Short Pulse Spectroscopy, Max-Born-Strasse 2A, D-12489 Berlin, Germany \\
 $^{2}$Institute of Physics, University of Rostock, Albert-Einstein-Str. 23, D-18059 Rostock, Germany
}

%


\begin{abstract}
The formation of an electron-hole plasma during the interaction of intense femtosecond
laser pulses with transparent solids lies at the heart of femtosecond laser processing. Advanced micro- and nanomachining applications require 
improved control over the excitation characteristics. Here, we relate the emission of 
low-order harmonics to the strong laser-field-induced plasma formation. Together with a measurement of the total plasma density we identify
the contribution of two competing ionization mechanisms - strong-field and electron-impact ionization. 
\end{abstract}

\maketitle
\section{Introduction}
The precise modification of solid targets with ultrashort laser pulses (USP)
has become a key technology in various fields ranging from data storage \cite{Glezer1995} to waveguide 
writing \cite{Davis1996, Miura1997, Mittholiya2017} and a multitude of cutting purposes \cite{Nisar2013, Ahmed2008}. Every permanent modification of the
optical properties of a target irradiated with USP is initiated by the formation of 
an electron-hole plasma \cite{Linde1996, Schaffer2001}. Motivated by the tremendous potential of ultrashort pulse
laser micro- and nanomachining \cite{Gattass2008, Mottay2016, Sugioka2014, Phillips2015, Malin2016, Furch2019}, petahertz electronics \cite{Garg2016, Yang2020, Goulielmakis2007, Schiffrin2013, Schultze2013, Sederberg2020} and coherent control \cite{Greenland2010, Horstmann2020}, huge 
theoretical and experimental effort has been
devoted to the description of laser-matter interactions at intensities around the
threshold for an irreversible laser-induced modification \cite{Lebugle2014, Quere2001, Mao2004, Winkler2006}. Although the main excitation 
mechanisms relevant for USP-driven laser material processing are unequivocally established as strong-field ionization (SFI) and electron-impact ionization (IMP), their relative importance
generally remains undetermined. Experimentally distinguishing the influence of SFI and IMP 
is rather challenging since both mechanisms lead to an intrapulse increase of the electron-hole plasma density \cite{Kaiser2000}. Indirect measurements performed to reveal the relative role of SFI
and IMP only allow the qualitative identification of the prevailing excitation mechanism \cite{Jupe2009, Gertsvolf2010}. \\
The population transfer provided by SFI not only depends on the intensity envelope of the
laser pulse but also on its sub-cycle structure \cite{Yudin2001, Zhokhov2018, Gertsvolf2010}. While femtosecond transient absorption 
and reflectivity measurements have emerged as reliable tools for the characterization of the cycle-averaged
plasma density buildup \cite{Audebert1994, Martin1997, Mero2003, Lechuga2017, Winkler2017, Moller2020}, sub-cycle carrier dynamics in solids have been studied by attosecond
transient absorption spectroscopy \cite{Schultze2013, Schultze2014, Schlaepfer2018} and attosecond polarization spectroscopy \cite{Sommer2016}. Due to the 
concomitant manifestation of other electronic processes such as the dynamical Franz-Keldysh
effect \cite{Lucchini2016, Otobe2016}, and IMP \cite{Kaiser2000}, these time-domain methods are not capable of isolating the contribution
of SFI to the ionization yield. \\
Several theoretical as well as experimental studies support the existence of a unique optical signature of SFI \cite{Brunel1990, Mitrofanov2011, Verhoef2010, Babushkin2017, Juergens2020, Li2020, Campi2021} in the frequency domain, which arises from the 
nearly stepwise increase of the plasma density $\rho_{\rm{SFI}}(t)$ twice per optical cycle of the  
driving laser field \cite{Siders2001, Serebryannikov2009}. However, the relative role of the involved physical processes in the associated frequency conversion process remained elusive. \\
In a recent publication we unequivocally demonstrated SFI-driven, non-perturbative wave-mixing in strongly exited fused silica below the damage threshold and identified the previously disregarded injection current as a dominant source of low-order harmonic generation \cite{Juergens2020}.
Here, we exploit the optical signature of this injection current to selectively extract the plasma formation induced by SFI. We demonstrate that the cycle-averaged conduction band electron density increase can be obtained from an analysis of the temporal signature of an arbitrary harmonic order. We further combine the time-resolved measurement of low-order, injection harmonics with time-domain ptychography \cite{Spangenberg2015, Witting2016} to access the sub-cycle dynamics of $\rho_{\rm{SFI}}(t)$.  \\
\\
\section{Experimental Setup}
The experimental apparatus is described in detail in Ref.~\cite{Juergens2020}. In brief, an electron-hole plasma was produced in a \SI{0.5}{\milli\meter}-thick a-SiO$_2$ sample (UV grade fused silica, Corning 7980) by focusing a short-wavelength infrared (SWIR, $\lambda_{\textrm{pump}} = $ \SI{2.1}{\micro\meter}) laser pulse along with a weak, time-delayed \SI{790}{\nano\meter}  (near-infrared, NIR) probe laser pulse using a cross-polarized, close-to-collinear pump-probe geometry. Employing cross-polarized pump and probe laser beams ensures that the weak probe laser pulse does not significantly modify the ionization yield (see Fig.~\ref{fig:para_cross_plasma}). Both laser pulses were obtained using a Ti:Sapphire regenerative amplifier (\SI{45}{\femto\second}, \SI{3.3}{\milli\joule} pulse energy, \SI{1}{\kilo\hertz} repetition rate, linearly polarized) that pumped an optical parametric amplifier (TOPAS C, Light Conversion). An off-axis parabolic mirror (focal length of \SI{50}{\milli\meter}, gold-coated) focused the SWIR beam down to a spot size of $\sim$\SI{35}{\micro\meter} ($1/e^2$ beam diameter in air) within the sample. These focusing conditions induced a local intensity in the focal region of up to \SI{15}{\tera\watt\per\centi\meter\squared}. The NIR probe beam was directed through a central hole in the parabolic mirror at a small angle ($<\,$\SI{5}{\degree}). The spectral properties of the light emerging from the sample were measured as a function of the pump-probe delay $\tau$ with the help of a commercial  visible / ultraviolet (VIS/UV) spectrometer (Avantes AvaSpec-HS1024x58/122TEC). Due to the noncollinear pump-probe geometry the two-color and the single-color wave-mixing signals propagated in different directions. The spectrometer was aligned such, that within its spectral range only the two-color, sum-frequency mixing signals were detected. \\
Simultaneously, the transmitted NIR probe laser pulse was analyzed with the help of a photodiode (Ophir PD-10) and an NIR spectrometer (Avantes AvaSpec-LS2048), in order to quantify the absorption by the plasma and modifications of the spectral properties, respectively. \\
\\
\section{Results}
\begin{figure}[htbp]
\centering
\includegraphics[width=\linewidth]{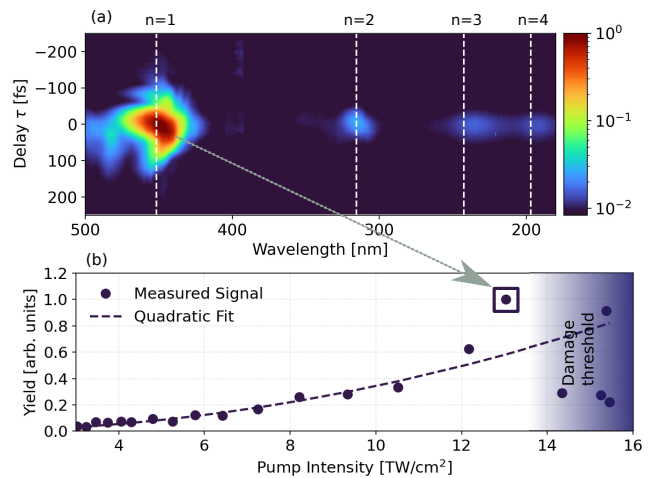}
\caption{Two-color wave-mixing in fused silica.  (a)  Time-frequency map of the two-color wave-mixing signal obtained by exciting the fused silica sample with a pump intensity of \SI{13}{\tera\watt\per\centi\meter\squared}. The vertical dashed lines indicate the positions of the first four harmonic orders predicted by Eq.~\ref{eq:harm_freqs}. (b) Yield of the $n=1$ wave-mixing signal as a function of the pump laser intensity. Figure~1(a) was measured at the point where the $n=1$ yield significantly deviates from the quadratic fit (dashed line).}
\label{fig:fig_1}
\end{figure}

A spectrogram displaying the emitted radiation as a function of the relative delay $\tau$ between both laser pulses is shown in Fig.~\ref{fig:fig_1}(a). An intense, transient, harmonic emission was observed at frequencies
\begin{equation}
\label{eq:harm_freqs}
\omega_n = 2n \times \omega_{\textrm{pump}} + \omega_{\textrm{probe}} \quad \text{with} \quad n = 1,2,3,4 
\end{equation}
(indicated by the vertical dashed lines).  
Figure~\ref{fig:fig_1}(b) displays the yield of the $n=1$ wave-mixing signal as a function of the pump laser intensity. At intensities $\leq $\SI{12}{\tera\watt\per\centi\meter\squared} the nearly quadratic dependence of the measured yield on the pump laser intensity indicates that the harmonic formation is dominated by Kerr-type four-wave-mixing, using two pump photons and one probe photon. The deviation from the parabolic dependence at pump laser intensities approaching the damage threshold of the SiO$_2$ substrate (i.e. at $\geq$ \SI{14}{\tera\watt\per\centi\meter\squared}) results from a change of the prevailing wave-mixing mechanism from a Kerr-type nonlinearity at low to intermediate intensity to SFI at high intensity.

Following Ref.~\cite{Juergens2019} the time-derivative of the current density associated with the strong-field-induced population transfer from the valence to the conduction band and the subsequent field-driven classical carrier dynamics in the conduction band is given by \cite{Geissler1999}
\begin{equation} \label{eq:geissler_current}
\frac{\partial \textbf{j}}{\partial t} = q_e n_0 \left[ \underbrace{\frac{q_e}{m_e} \textbf{E} \rho_{\rm{SFI}}}_{\rm{Brunel}} +\underbrace{\textbf{v}_0\dot{\rho}_{\rm{SFI}}}_{\rm{velocity}} + \underbrace{\frac{\partial}{\partial t} (\textbf{x}_0 \dot{\rho}_{\rm{SFI}})}_{\rm{injection}} \right]
\end{equation}
where $q_e = -e$ is the electron charge, $n_0$ is the SiO$_2$ molecular density, $m_e$ is the electronic mass, $\textbf{E}$, the driving laser field, $\rho_{\rm{SFI}}$ the density of electrons in the conduction band and $\dot{\rho_{\rm{SFI}}}$ the SFI rate. The parameters $\textbf{x}_0$ and $\textbf{v}_0$ denote the spatial displacement and the velocity of the electrons after injection into the conduction band. Regarding the associated contributions to nonlinear wave-mixing, it has been demonstrated that the spatial displacement of electrons during the transition from the valence to the conduction band [in other words the injection current $j_{\rm{inj}}(t)$] dominates over the Brunel current and the velocity-current over a wide range of parameters. In a cross-polarized pump-probe geometry where the polarization of the harmonic emission follows the polarization of the probe laser pulse and only a single probe photon is used in the frequency conversion process, the injection current can be simplified to
\begin{align} 
\label{eq:injection_current}
j_{\rm{inj}}(t, \tau) &= \frac{n_0 E_g \dot{\rho}_{\rm{SFI}}(t)}{E_{\rm{pump}}(t)^2} E_{\rm{probe}}(t-\tau) \notag \\
&\equiv \sigma_{\rm{inj}}(t) E_{\rm{probe}}(t-\tau).
\end{align}
For convenience we have introduced the effective conductivity 
\begin{align}
\label{eq:sigma}
\sigma_{\rm{inj}}(t) &= \frac{n_0 E_g \dot{\rho}_{\rm{SFI}}(t)}{E_{\rm{pump}}(t)^2} \notag \\
& \propto \frac{\dot{\rho}_{\rm{SFI}}(t)}{E_{\rm{pump}}(t)^2} \notag \\
& \approx \frac{\dot{\rho}_{\rm{SFI}}(t)}{I_{\rm{pump}}(t)}
\end{align}
as the quantity that will be retrieved from a measurement of the injection harmonics as a function of the pump-probe delay. It should be emphasized that replacing $E_{\rm{pump}}(t)^2$ by the cycle-averaged intensity of the SWIR pump laser pulse $I_{\rm{pump}}(t)$ generally constitutes a strong approximation whose justification has to be checked carefully. Here, however, it is applicable due to the strongly localized character and the high degree of nonlinearity of $\dot{\rho}_{\rm{SFI}}(t)$ (see supplemental document). As the effective, strong-field-induced conductivity is nonzero only near the crests of individual laser half cycles, the field in the denominator can effectively be replaced by its amplitude without introducing a substantial error. \\
 In the next step, the SFI rate $\dot{\rho}_{\rm{SFI}}(t)$ can be decomposed into the sum of a smooth, cycle-averaged component - $\dot{\rho}_{\rm{SFI}}^{\rm{ca}}(t)$ - and an oscillating, sub-cycle component - $\dot{\rho}_{\rm{SFI}}^{\rm{sub}}(t)$ (for details see supplemental document). Only the sub-cycle plasma formation due to SFI leads to the harmonic emission that is reported in Fig.~\ref{fig:fig_1}(a). Finally, the experimentally measured harmonic yield $Y_{\rm{inj}}(\omega, \tau)$ can be related to the sub-cycle ionization dynamics according to (for a derivation see supplemental document):
 \begin{align}
 \label{eq:measured_quantity}
 Y_{\rm{inj}}(\omega, \tau) & \propto \left\vert \int \frac{\partial j_{\rm{inj}}^{\rm{sub}}(t,\tau)}{\partial t} e^{i \omega t} dt \right\vert ^2 \notag \\
 & = \omega^2 \left\vert \int \sigma_{\rm{inj}}^{\rm{sub}}(t) E_{\rm{probe}}(t-\tau) e^{i \omega t} dt \right\vert^2
\end{align}
%
\subsection{Cycle-averaged ionization dynamics}
\begin{figure}[htbp]
\centering
\includegraphics[width=\linewidth]{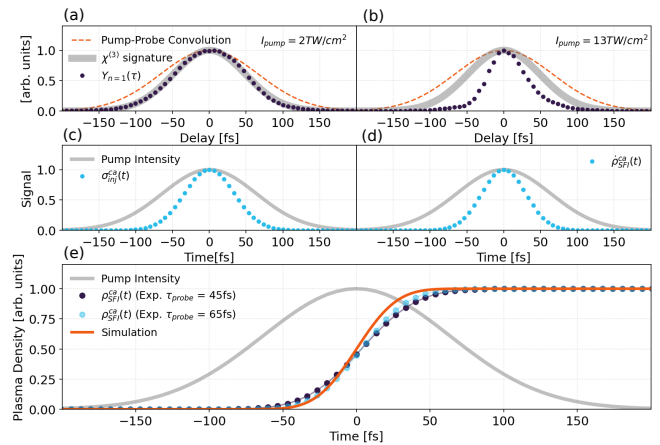}
\caption{Linking injection harmonics to cycle-averaged carrier dynamics by a direct deconvolution. (a) Intensity $Y_{n=1}(\tau)$ of the $n=1$ harmonic as a function of the pump-probe delay obtained at a pump intensity of \SI{2}{\tera\watt\per\centi\meter\squared}, together with the pump-probe convolution and the corresponding temporal profile of a $\chi^{(3)}$-driven wave-mixing process. (b) Same as in (a) for a pump intensity of \SI{13}{\tera\watt\per\centi\meter\squared}. (c) Effective, strong-field-induced conductivity $\sigma_{\rm{inj}}(t)$, derived from (b), together with the intensity envelope of the pump laser pulse. (d) Cycle-averaged SFI rate obtained by multiplying $\sigma_{\rm{inj}}(t)$ with $I_{\rm{pump}}(t)$. (e) Plasma density induced by SFI, resulting from a direct integration of the SFI rate shown in (d), compared to a numerical result for $\rho_{\rm{SFI}}^{\rm{ca}}(t)$.}
\label{fig:cycle_averaged}
\end{figure}

In order to demonstrate the link of the emission of injection harmonics to the strong-field-induced plasma formation we focus on the $n=1$ harmonic. Figure \ref{fig:cycle_averaged}(a) and (b) show the experimental intensity of the $n=1$ harmonic, $Y_{n=1}(\tau)$, as a function of the pump-probe delay, compared to the convolution of the pump and probe laser pulses and the expected signature of a $\chi^{(3)}$ wave-mixing process ($I_{\rm{probe}}(t-\tau) \otimes I_{\rm{pump}}(t)^2$, with linear convolution operator $\otimes$) at different pump laser intensities. At $I_{\rm{pump}} =$ \SI{2}{\tera\watt\per\centi\meter\squared}, where four-wave-mixing is assumed to dominate [see Fig.~\ref{fig:fig_1}(b)], the spectrally integrated $n=1$ harmonic signal agrees very well with the $\chi^{(3)}$ signature while at $I_{\rm{pump}} =$ \SI{13}{\tera\watt\per\centi\meter\squared} the emission is temporally more confined, as a consequence of the significantly higher order nonlinearity of the generation process. \\ 
At near-threshold intensities, the dependence of the harmonic intensity on the pump-probe delay is related to a convolution of the effective conductivity with the probe laser pulse. Analogous to the plasma density and the SFI rate, the $\sigma_{\rm{inj}}$ can be divided into a cycle-averaged and a sub-cycle component. Only the cycle-averaged dynamics of the sub-cycle modulated, effective conductivity can be extracted from the delay dependence of the harmonic intensity that is given by $Y_{\rm{inj}}^{n=1}(\tau) \propto \left\vert \sigma_{\rm{inj}}^{\rm{ca}}(t) \otimes E_{\rm{probe}}(t-\tau) \right\vert^2$. By means of a numerical deconvolution using the temporal profile of the NIR probe laser pulse, the cycle-averaged contribution of $\sigma_{\rm{inj}}(t)$ can be retrieved [see Fig.~\ref{fig:cycle_averaged}(c)]. Multiplication of this conductivity with the intensity envelope of the pump laser pulse $I_{\rm{pump}}(t)$ (see Eq.~\ref{eq:injection_current}) reveals the cycle-averaged SFI rate - $\dot{\rho}_{\rm{SFI}}^{\rm{ca}}(t)$ - shown in Fig.~\ref{fig:cycle_averaged}(d). The cycle-averaged, strong-field-induced plasma density buildup can be computed by a direct integration of $\dot{\rho}_{\rm{SFI}}^{\rm{ca}}(t)$ and is compared to a numerical simulation under our experimental conditions in Fig.~\ref{fig:cycle_averaged}(e) (the used numerical model is described in detail in Ref.~\cite{Juergens2020}). The plasma density obtained from the experiments starts to increase at slightly earlier times with a slightly lower slope than the numerically determined one. A possible explanation for the deviation between the numerical prediction and the experimentally determined plasma formation is an uncertainty of the temporal profile of the probe laser pulse in the interaction region. The retrieval of $\rho_{\rm{SFI}}^{\rm{ca}}(t)$ from a direct deconvolution of $Y_{\rm{inj}}^{n=1}(\tau)$ requires the knowledge of the temporal envelope of $E_{\rm{probe}}(t)$. Due to propagation through the fused silica sample and the electron-hole plasma the temporal properties of the probe laser pulse may be modified leading to an additional uncertainty in the retrieval of the plasma density buildup. Note that the described characterization of the cycle-averaged plasma dynamics can be performed using an arbitrary harmonic order, as all injection harmonics exhibit an identical temporal signature (for details see supplemental document). The analysis of higher orders is particularly attractive as they are quasi background-free with respect to Kerr-type contributions. \\
%
\\
\subsection{Sub-cycle ionization dynamics}
The previous analysis of the time-resolved injection harmonics connected the delay-dependent intensity of the $n=1$ harmonic to the cycle-averaged plasma dynamics induced by SFI. In what follows, a more robust numerical method is introduced that does not require an a priori knowledge of the probe laser field and furthermore gives insight into the sub-cycle dynamics of the SFI-induced plasma formation. \\
Equation \ref{eq:measured_quantity} contains the product of two complex-valued functions whose phase information is lost as the measured quantity $Y_{\rm{inj}}(\omega, \tau)$ is a purely real-valued object. Hence, Eq.~\ref{eq:measured_quantity} resembles a classical phase-retrieval problem that is well-known in nonlinear optics from many FROG/X-FROG applications (see e.g. \cite{Trebino1993, Trebino1997, Linde1998}). \\
In order to deduce information on the sub-cycle dynamics of the underlying SFI process we numerically retrieve $\sigma_{\rm{inj}}^{\rm{sub}}(t)$ by a ptychographic analysis of $Y_{\rm{inj}}(\omega, \tau) / \omega^2$. Multiplication of the retrieved effective conductivity with the intensity envelope of the pump laser pulse $I_{\rm{pump}}(t)$ reveals the oscillatory dynamics of the SFI rate [i.e. $\dot{\rho}_{\rm{SFI}}^{\rm{sub}}(t)$]. To reconstruct the full SFI rate, we analyze the temporal envelope of $\rho_{\rm{SFI}}^{\rm{sub}}(t)$. As detailed in the supplemental document, the envelope of the retrieved sub-cycle component of the SFI rate corresponds to the cycle-averaged SFI rate $\dot{\rho}_{\rm{SFI}}^{\rm{ca}}(t)$. Adding up the two components while fulfilling the condition $\dot{\rho}_{\rm{SFI}}(t) \geq 0 \, \forall \, t$ (i.e. $\dot{\rho}_{\rm{SFI}}(t) = \alpha \dot{\rho}_{\rm{SFI}}^{\rm{ca}}(t) + \dot{\rho}_{\rm{SFI}}^{\rm{sub}}(t)$ with $\alpha = \vert \textrm{min}  \{ \dot{\rho}_{\rm{SFI}}^{\rm{sub}}(t)\} \vert$) yields the full SFI rate that can be directly integrated to retrieve the plasma formation induced by SFI from the ptychographic analysis of time-resolved, low-order harmonic spectra. \\
\begin{figure*}[htbp]
\centering
\includegraphics[width=\linewidth]{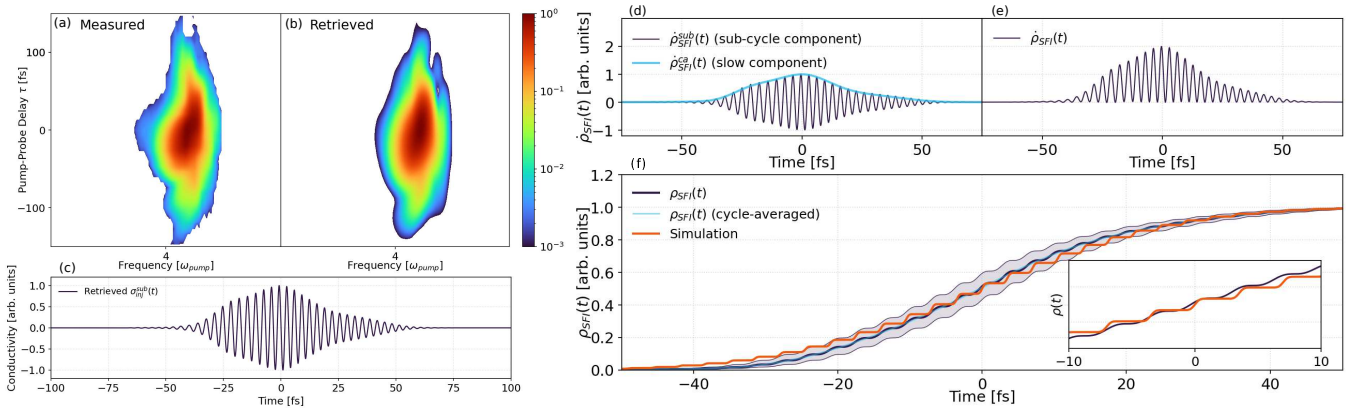}
\caption{Relating the time-resolved measurement of injection harmonics to sub-cycle ionization dynamics . (a). Measured time-frequency map of the two-color wave-mixing signal at an SWIR excitation intensity of \SI{13}{\tera\watt\per\centi\meter\squared}. (b). Retrieved spectrogram by the ptychographic iterative engine after 1000 iterations. (c). Effective, strong-field-induced conductivity $\sigma_{\rm{inj}}(t)$ as retrieved by time-domain ptychography. (d). Reconstructed fast and slow component of the SFI rate - $\dot{\rho}_{\rm{SFI}}(t)$. (e). Resulting total ionization rate due to SFI obtained from experimental, two-color, harmonic spectra. (f). Strong-field-induced plasma formation as retrieved from the time-resolved measurement of the $n=1$ harmonic compared to numerical simulations under our experimental conditions.}  
\label{fig:fig_3}
\end{figure*}
%
A combined analysis of the slow (femtosecond timescale in the current experiment) dynamics that are encoded in the delay dependence of the harmonic intensity and the spectral properties of the emitted harmonics in principle allows a full reconstruction of the sub-cycle ionization dynamics induced by the strong SWIR laser field. However, doing so requires a full (amplitude \& phase) retrieval of the harmonics. Since the neighbouring harmonic orders do not spectrally overlap [see Fig.~\ref{fig:fig_1}(a)], the relative phase between two consecutive harmonics remains unknown and we continue our analysis with the $n=1$ harmonic only.\\
Results of the ptychographic analysis of the measured $n=1$ harmonic are displayed in Fig.~\ref{fig:fig_3}. Figure~\ref{fig:fig_3}(a) and (b) show the measured and retrieved time-frequency map containing the $n=1$ harmonic. The ptychographic algorithm is able to reconstruct the measured wave-mixing signal with high accuracy down to the noise level of the experiment. The sub-cycle-modulated, effective conductivity $\sigma_{\rm{inj}}^{\rm{sub}}(t)$ retrieved by the ptychographic algorithm is shown in Fig.~\ref{fig:fig_3}(c). The resulting sub-cycle component of the SFI rate - $\dot{\rho}_{\rm{SFI}}^{\rm{sub}}(t)$ is depicted together with the cycle-averaged, low-frequency contribution in Fig.~\ref{fig:fig_3}(d) (Note that due to the decomposition $\dot{\rho}_{\rm{SFI}}(t) = \dot{\rho}_{\rm{SFI}}^{\rm{ca}}(t) + \dot{\rho}_{\rm{SFI}}^{\rm{sub}}(t)$ the sub-cycle component of the SFI rate can have negative values, while the cycle-averaged and the total SFI rate are always $\geq 0$). Combining the fast and the slow component leads to the total SFI rate depicted in Fig.~\ref{fig:fig_3}(e). Direct integration of $\dot{\rho}_{\rm{SFI}}(t)$ yields the total plasma formation due to SFI reconstructed from the time-resolved measurement of the $n=1$ harmonic. The resulting $\rho_{\rm{SFI}}(t)$ is shown together with results obtained by numerical simulations under identical conditions in Fig.~\ref{fig:fig_3}(f). A nearly stepwise increase of the resulting plasma density with two steps per optical cycle of the SWIR driving field is found. As the carrier-envelope phase (CEP) of the pump laser pulses is not actively stabilized in our experiments and the ptychographic algorithm is not able to retrieve the absolute phase of $\sigma_{\rm{inj}}^{\rm{sub}}(t)$ our results exhibit a temporal uncertainty of one optical cycle [see shaded area in Fig.~\ref{fig:fig_3}(f)]. \\
When compared to the numerically obtained plasma density buildup, a slightly later liftoff of the $\rho_{\rm{SFI}}(t)$ obtained from our measurement is observed. The overall slope of the reconstructed $\rho_{\rm{SFI}}(t)$ is found to be marginally steeper than the numerical prediction. Discrepancies between the reconstructed and numerically predicted $\rho_{\rm{SFI}}(t)$ might be due to dynamic modifications of the bandgap \cite{Winkler2017, Lucchini2016} leading to transiently enhanced or diminished ionization probabilities. \\
\\
\\
\section{Discussion and Outlook}
\begin{figure}[htbp]
\centering
\includegraphics[width=\linewidth]{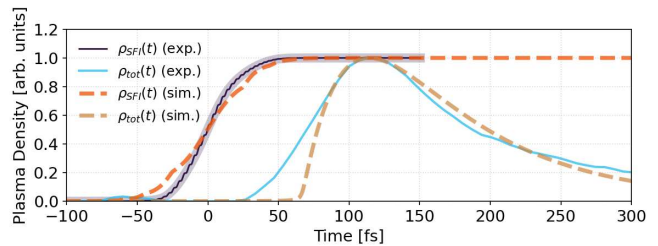}
\caption{Comparison of the SFI-induced plasma dynamics extracted from the ptychographic analysis of the $n=1$ harmonic and the total plasma dynamics obtained from a time-resolved absorption measurement. The dashed lines show numerical results for $\rho_{\rm{SFI}}(t)$ and $\rho_{\rm{tot}}(t)$ at a pump intensity of \SI{17}{\tera\watt\per\centi\meter\squared}.}
\label{fig:fig_4}
\end{figure}

While the cycle-averaged dynamics can be extracted from the temporal signature of an arbitrary harmonic order (here, we chose the $n=1$ harmonic), the stepwise increase of the plasma density can be reconstructed by a combined analysis of the injection harmonics in the time and frequency domain. The $n=1$ harmonic suffices to characterize the cycle-averaged plasma dynamics and additionally provides valuable insights into the sub-cycle dynamics. As a consequence, it is possible to apply the ptychographic reconstruction of strong-field-induced carrier dynamics on a wide range of materials including semiconductors with bandgaps in the visible spectral range where only $\hbar \omega_{n=1} \leq E_g $. \\
In addition to the retrieval of $\sigma_{\rm{inj}}^{\rm{sub}}(t)$ time-domain ptychography retrieves the temporal profile of the NIR probe laser pulse in the interaction region. The retrieved duration ($\sim$\SI{65}{\femto\second}) differs significantly from the initial duration (\SI{45}{\femto\second}) indicating that the propagation through the fused silica sample and the electron-hole plasma strongly affect the temporal structure of the probe pulse. Indeed, using the probe laser pulse retrieved by the ptychographic iterative engine for the deconvolution considerably improves the agreement between the retrieved and the simulated $\rho_{\rm{SFI}}^{\rm{ca}}(t)$ [see Fig.~\ref{fig:cycle_averaged}(e)].
\\
Finally, we are able to disentangle the plasma dynamics induced by SFI from the competing ionization and relaxation mechanisms. Together with the simultaneously measured transmission of the NIR probe laser that carries information on the total plasma density (see Ref.~\cite{Juergens2019}) this enables the determination of the relative contribution of SFI and IMP. \\
Figure~\ref{fig:fig_4} shows a comparison between the total plasma density obtained from a time-resolved transmission measurement and the SFI-induced plasma formation extracted from the time-resolved detection of the $n=1$ injection harmonic. Here, both densities are normalized to unity as the absolute number of excited carriers cannot be determined by both presented methods. When comparing the temporal profile of $\rho_{\rm{tot}}(t)$ and $\rho_{\rm{SFI}}(t)$ it becomes evident that $\rho_{\rm{SFI}}(t)$ starts to increase at much earlier times (at $\tau\approx$ \SI{-50}{\femto\second}). $\rho_{\rm{tot}}(t)$ lifts off at $\tau \approx$ \SI{25}{\femto\second} where $\rho_{\rm{SFI}}(t)$ already reaches $90\%$ of its maximum value.  The fact that the experimentally determined $\rho_{\rm{SFI}}(t)$ reaches its maximum value where no significant probe absorption occurs indicates that IMP is the dominating excitation mechanism under our experimental parameters. This is further supported by the delay between the peak of the pulse and the maximum of $\rho_{\rm{tot}}(t)$ at $\approx$ \SI{110}{\femto\second} and is in very good agreement with numerical simulations that predict a complete dominance of IMP for our experimental configuration. At pump-probe delays $\geq$ \SI{110}{\femto\second}, $\rho_{\rm{tot}}(t)$ exhibits an exponential decay due to ultrafast relaxation processes that are not accounted for in the determination of $\rho_{\rm{SFI}}(t)$. Therefore, $\rho_{\rm{SFI}}(t)$ stays at the maximum level for longer time delays.\\
\\
%
In conclusion, we have presented time-resolved experiments on two-color, low-order harmonic generation in fused silica with the aim of isolating the plasma density buildup induced by SFI. Our analysis demonstrates a possible application of the recently identified injection harmonics for ultrafast metrology in solid state systems. The fact that the detected signal is not screened by concomitant ionization mechanisms such as IMP, allows a background-free investigation of strong-field phenomena in bulk solids. \\
We demonstrated the possibility to extract the cycle-averaged plasma density buildup due to SFI from a simple numerical analysis of the intensity of an arbitrary harmonic order. By applying an iterative phase retrieval algorithm on the time-resolved measurement of the $n=1$ harmonic we revealed the stepwise character of the field-induced plasma formation. The exact shape of these steps can be obtained by including higher harmonic orders with a well-determined relative phase while the exact temporal position of the ionization bursts is determined by the relative phase between the driving laser field and the emitted harmonics. Comparing the SFI-induced plasma formation to the total evolution of the electron-hole plasma provides intimate insights on the relative contribution of field-driven and impact-driven ionization channels to the total plasma density buildup. Importantly, SFI-induced signals manifest well-below the threshold for a permanent modification and thus hold the potential for an in-situ detection of the onset of optical damage before a catastrophic damage occurs.

\subsection*{Funding} 
Deutsche Forschungsgemeinschaft (IDs DFG ME4427/1-1 and DFG ME4427/1-2, Heisenberg Grant ID 398382624, SPP1840: QUTIF ID 281272685), Leibniz Association grant SAW‐LAPTON (K266/2019), Ministry of Education, Science and Culture of Mecklenburg-Vorpommern, Germany via the project "NEISS".
\subsection*{Acknowledgements}
We thank M. Jup{\'e} (Laser Zentrum Hannover) for measuring the bandgap of the samples used in this study. 
\subsection*{Supplemental document}
See supplemental document for supporting content.

\bibliography{PIE_arxiv}
\newpage 
\section*{Supplemental Document}
\label{sec:appendix}
\renewcommand{\thefigure}{S\arabic{figure}}
\setcounter{figure}{0}

\subsection{Two-color plasma formation using parallel and perpendicular polarizations}
\begin{figure}[htbp]
\centering
\includegraphics[width=\linewidth]{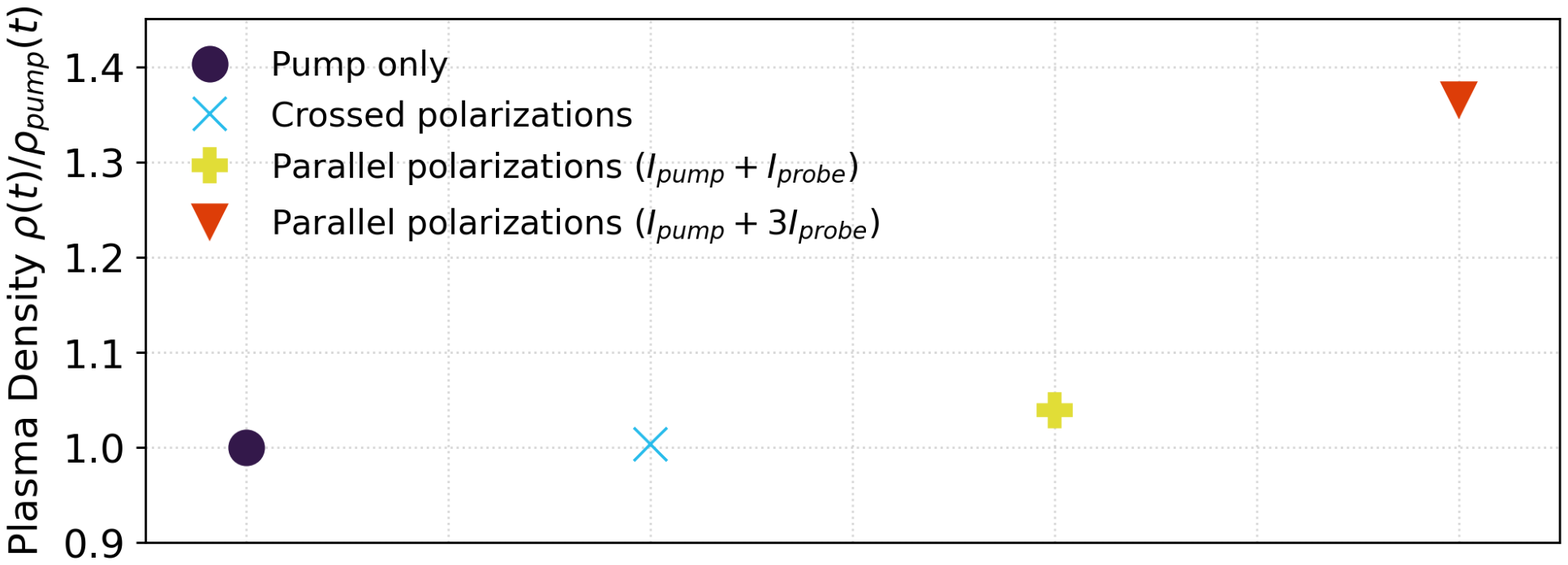}
\caption{Calculated strong-field-induced plasma dynamics due to two-color excitation of fused silica in different polarization configurations normalized to the yield obtained from irradiation with the pump laser pulse only. Simulation parameters: $I_{\rm{pump}} = $~\SI{12}{\tera\watt\per\centi\meter\squared}, $\lambda_{\rm{pump}} = $~\SI{2100}{\nano\meter}, $\tau_{\rm{pump}} = $~\SI{140}{\femto\second}, $I_{\rm{probe}} = $~\SI{0.015}{\tera\watt\per\centi\meter\squared}, $\lambda_{\rm{probe}} = $~\SI{800}{\nano\meter}, $\tau_{\rm{probe}} = $~\SI{45}{\femto\second}.}
\label{fig:para_cross_plasma}
\end{figure}

Numerical simulations of the plasma formation due to SFI using the model described in \cite{Juergens2020} illustrate the importance of the perpendicular polarizations of the pump and probe laser fields. Figure~\ref{fig:para_cross_plasma} shows the number of electrons promoted into the conduction band by the pump laser field only, the pump laser field in combination with a weak, cross-polarized probe laser field, the pump laser field and a weak parallel-polarized probe laser field ($I_{\rm{pump}} + I_{\rm{probe}}$ and the pump field and a probe laser field of thrice the initial intensity ($I_{\rm{pump}} + I_{\rm{probe}}$). While the ionization yield is essentially the same in the first two cases, in a parallel polarization configuration even a weak probe laser field is sufficient to substantially modify the total number of conduction band electrons generated during pump-probe overlap. Slight changes of the probe laser intensity considerably enhance the cross-excitation efficiency leading to a clear departure of the resulting plasma dynamics from the pump-field-induced plasma formation.

\subsection{Derivation of $Y_{\rm{inj}}(\omega, \tau)$}
Only the sub-cycle dynamics of the SFI rate contribute to the formation of the harmonics. The sub-cycle component of the injection current relevant for the emission of the $n=1$ wave-mixing signal, assuming cross-polarized pump and probe laser fields and a weak probe pulse can be expressed as
\begin{equation}
j_{\rm{inj}}^{\rm{sub}}(t,\tau) = \frac{n_0 E_g \dot{\rho}_{\rm{SFI}}^{\rm{sub}}(t)}{E_{\rm{pump}}(t)^2} E_{\rm{probe}}(t-\tau)
\end{equation}
The expression for the measured quantity $Y_{\rm{inj}}(\omega, \tau)$ can be derived by 
\begin{align}
    Y_{\rm{inj}}(\omega, \tau) &\propto \left\vert \int \frac{\partial j_{\rm{inj}}^{\rm{sub}}(t,\tau)}{\partial t} e^{i \omega t} dt \right\vert ^2 \notag \\
    & = \omega^2 \left\vert \int j_{\rm{inj}}^{\rm{sub}}(t, \tau) e^{i \omega t} dt \right\vert ^2 \notag \\
    & = \omega ^2 \left\vert \int \frac{n_0 E_g \dot{\rho}^{\rm{sub}}_{\rm{SFI}}(t)}{E_{\rm{pump}}(t)^2} E_{\rm{probe}}(t-\tau) e^{i \omega t} dt \right\vert ^2 \notag \\
    & = \omega^2 \left\vert \int \sigma_{\rm{inj}}^{\rm{sub}}(t) E_{\rm{probe}}(t-\tau) e^{i \omega t} dt \right\vert ^2
\end{align}
All variables are defined in the main text.
\subsection{Simplifications in order to analyze the retrieved quantity}
The calculation of $\dot{\rho}_{\rm{SFI}}(t)$ from Eq.~\ref{eq:measured_quantity} requires the exact knowledge of the pump laser field together with the relative phase between $E_{\rm{pump}}(t)$ and $\sigma_{\rm{inj}}(t)$, which is not accessible in the presented experiment. However, we can exploit the fact that the pump laser field varies slowly compared to the SFI rate due to its high nonlinearity. Since SFI takes place mainly close to the extrema of $E_{\rm{pump}}(t)$ it is possible to replace $E_{\rm{pump}}(t)^2$ by the intensity envelope $I_{\rm{pump}}(t)$ of the SWIR pump laser pulse.
This approximation is however only valid for certain experimental parameters. In our case, the high nonlinearity of the excitation process due to the small photon energy of the SWIR pump laser pulse (\SI{0.59}{\electronvolt}) in combination with the wide bandgap of the fused silica samples (\SI{7.7}{\electronvolt}) justifies the simplification. For shorter wavelengths (larger photon energies) or materials with a smaller bandgap this approximation tends to fail as the excitation process is temporarily less confined due to the lower nonlinearity of the SFI process. \\
%
\begin{figure}[htbp]
\centering
\includegraphics[width=\linewidth]{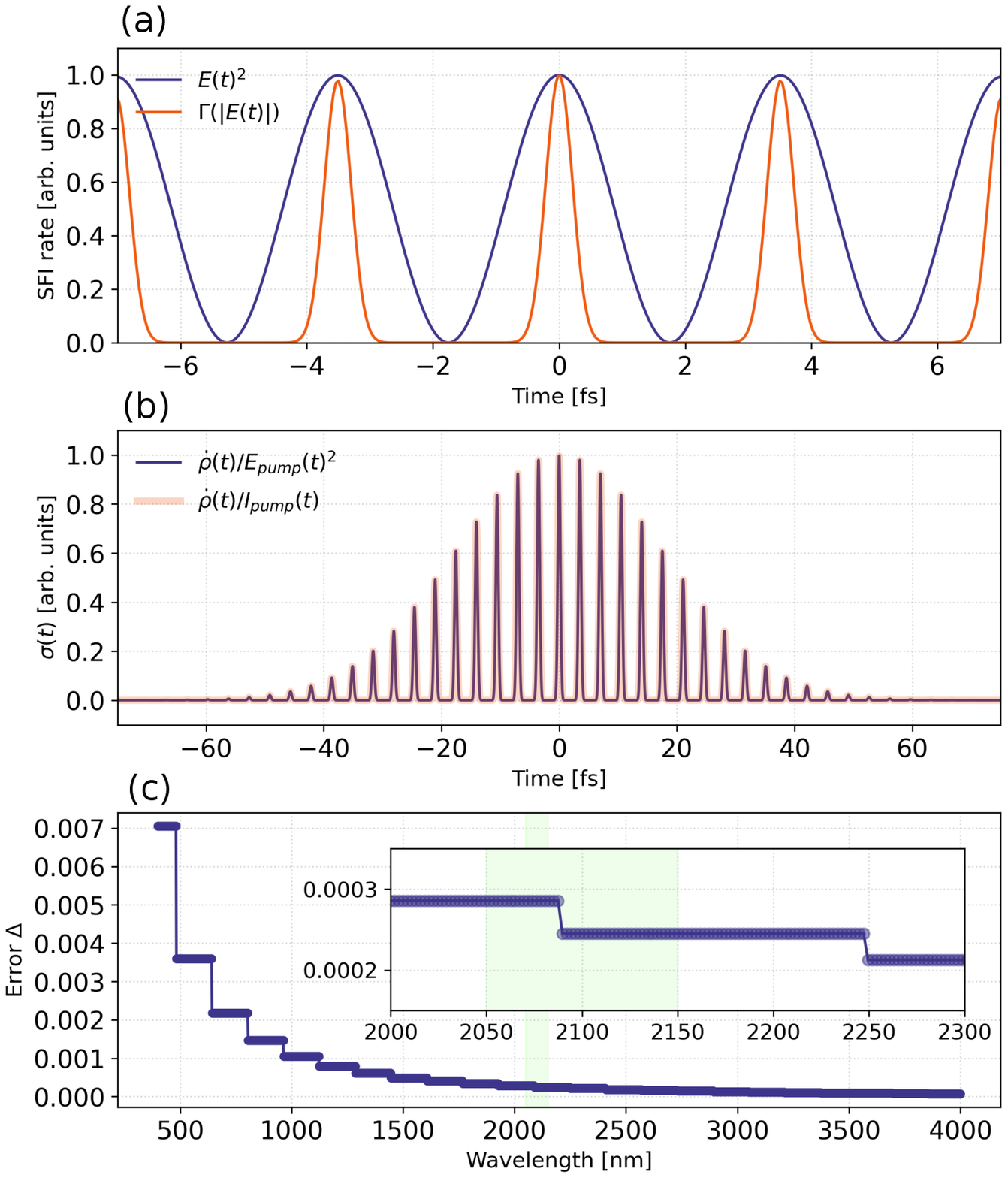}
\caption{Numerical analysis of the assumptions made for $\sigma_{\rm{inj}}(t)$ in the main text. (c) Comparison of the SFI rate and $E(t)^2$. (b). Normalized quantity $\dot{\rho}_{\rm{SFI}}(t) / E_{\rm{pump}}(t)^2$ compared to the discussed approximation $\dot{\rho}_{\rm{SFI}}(t) / I_{\rm{pump}}(t)$ and the ionization rate. (c) Error $\Delta$ resulting from the discussed simplification as a function of the wavelength.}
\label{fig:simple}
\end{figure}
Figure~\ref{fig:simple} illustrates this by comparing the SFI rate to the instantaneous electric field [see Fig.~\ref{fig:simple}(a)]. Due to the high nonlinearity of the SFI process, ionization takes place only in the vicinity of the peaks of $E(t)^2$ where $E(t)^2 \approx I(t)$. Figure~\ref{fig:simple}(b) compares the quantities $\sigma_{\rm{inj}}(t) = \dot{\rho}_{\rm{SFI}}(t) / E_{\rm{pump}}(t)^2$ to the simplification $\dot{\rho}_{\rm{SFI}}(t) / I_{\rm{pump}}(t)$ for the experimental parameters ($\lambda_{\rm{pump}} =\,$\SI{2.1}{\micro\meter}, $I_{\rm{pump}} = \,$ \SI{13}{\tera\watt\per\centi\meter\squared}). No difference between the exact conductivity and the approximation is observed. In order to analyze the influence of the wavelength on the validity of the approximation we can assume a pure multi-photon ionization scenario and approximate the ionization rate by $\dot{\rho}_{\rm{SFI}}(t) \propto E_{\rm{pump}}(t)^{2N}$ where $N$ denotes the multi-photon order defined by $N = ]E_{\rm{g}}/\hbar \omega + 1 [$. Here $E_{\rm{g}}$ is the bandgap of the material and $]x[$ denotes the integer part of $x$. The mean absolute error $\Delta$ shown in Fig.~\ref{fig:simple}(b) is calculated as follows:
\begin{equation}
\Delta = \frac{1}{M} \sum_{i=0}^{M} \left\vert \frac{E_{\rm{pump}}^{2 N}(t_i)}{E_{\rm{pump}}^{2}(t_i)} - \frac{E_{\rm{pump}}^{2N}(t_i)}{I_{\rm{pump}}(t_i)} \right\vert.
\end{equation}
The observed dependence of $\Delta$ on the excitation wavelength indicates that the replacement of $E_{\rm{pump}}(t)^2$ by the intensity envelope $I_{\rm{pump}}(t)$ is only valid for long wavelengths (high multi-photon order) while for shorter wavelengths the approximation is expected to substantially modify the result.

\subsection{Decomposition of the SFI rate}
\begin{figure}[htbp]
\centering
\includegraphics[width=\linewidth]{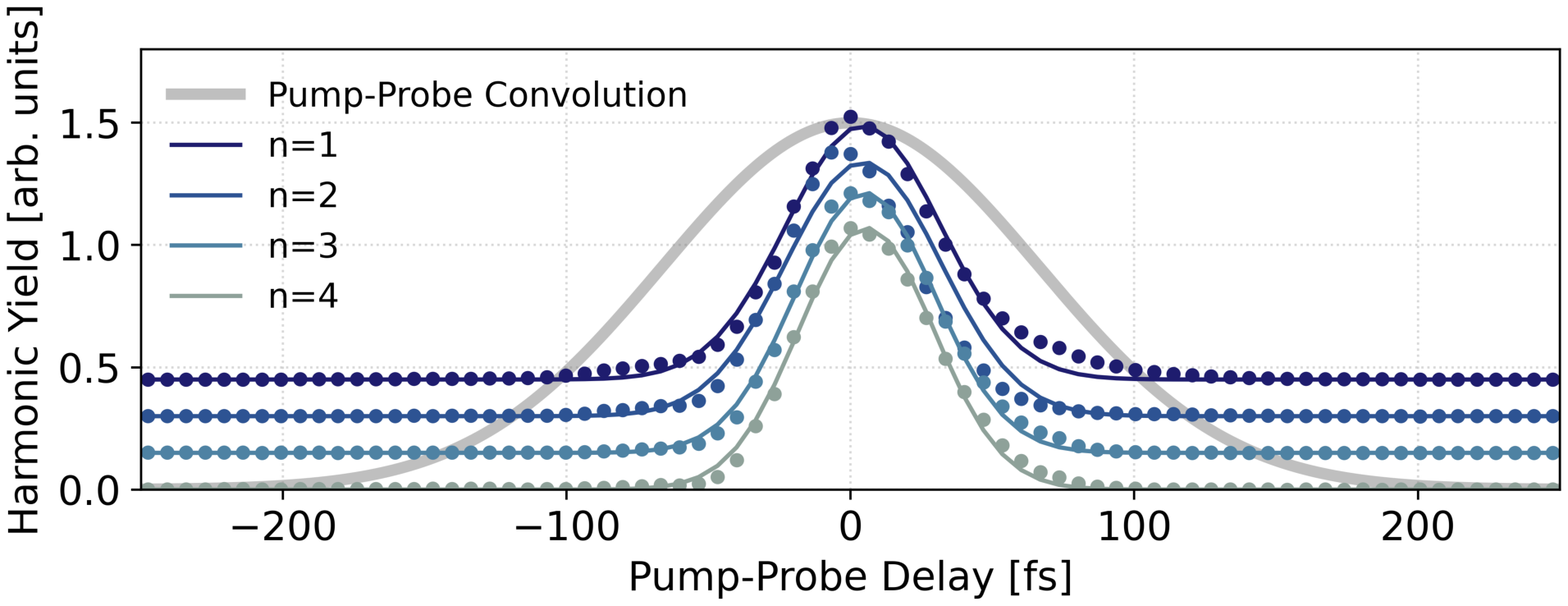}
\caption{Projection of the harmonic peaks from Fig.~\ref{fig:fig_1}(b) on the pump-probe delay axis compared to the pump-probe convolution.}
\label{fig:delay_projections}
\end{figure}

Assuming an intensity profile that is given by the product of an envelope and an oscillating part $I(t)=A(t) \cos(\omega_0 t)^2$, with amplitude $A(t)$ and central angular frequency $\omega_0$, the strong-field ionization rate $\dot{\rho}_{\rm{SFI}}(t)$ can be decomposed as a Fourier cosine series with a fundamental frequency of $2 \omega_0$ (see also \cite{Mermillod2019}):
\begin{figure}[htbp]
\centering
\includegraphics[width=0.8\linewidth]{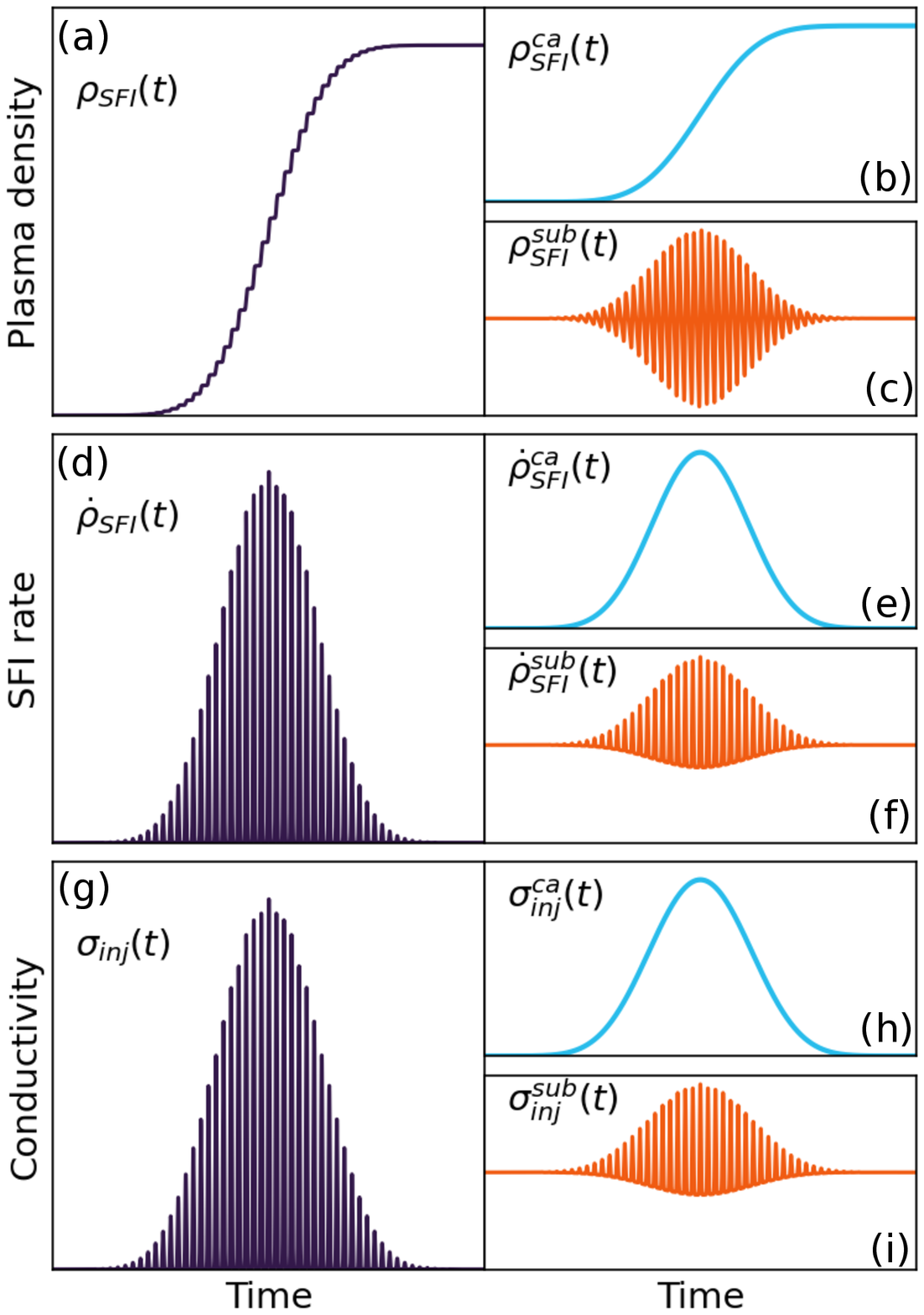}
\caption{Decomposition of the strong-field-induced plasma density, the SFI rate and the effective conductivity into a cycle-averaged and a sub-cycle component. (a)-(c) Strong-field-induced plasma formation - $\rho_{\rm{SFI}}(t)$. (d)-(f) SFI rate - $\dot{\rho}_{\rm{SFI}}(t)$. (g)-(i) Strong-field-induced conductivity - $\sigma_{\rm{inj}}(t)$.}
\label{fig:decomposition}
\end{figure}
\begin{align} \label{eq:series}
\frac{\partial \rho_{\rm{SFI}}(t)}{\partial t} &= B(t) \left [ \frac{1}{2} a_0 + \sum_{n=1}^{\infty} a_n(t) \cos(2 n \omega_0 t) \right ] \notag \\
&= \underbrace{\frac{a_0 B(t)}{2}}_{\rm{cycle-averaged}} + \underbrace{B(t) \sum_{n=1}^{\infty} a_n(t) \cos(2 n \omega_0 t)}_{\rm{sub-cycle}} 
\end{align}  
with $B(t)$ being a nonlinear function in $A(t)$ taking into account the intensity dependence of the SFI rate. Here $n$ denotes the order of the wave-mixing process and $a_i$ the $i$-th Fourier coefficient. Hence, the total plasma density induced by SFI is given by:
\begin{align} \label{eq:decom_rho}
\rho_{\rm{SFI}}(t) &= a_0 \int_{-\infty}^{t} \frac{B(t')}{2} dt' + \int_{-\infty}^{t} \frac{B(t')}{2} \sum_{n=1}^{\infty} a_n(t) \cos(2 n \omega_0 t') dt' \notag \\
& \approx \underbrace{\int_{-\infty}^{t} \frac{B(t')}{2} dt'}_{\rm{cycle-averaged}} + \underbrace{\frac{B(t')}{2} \sum_{n=1}^{\infty} \int_{-\infty}^{t} a_n(t) \cos(2 n \omega_0 t') dt'}_{\rm{sub-cycle}} .
\end{align}
The decomposition of the plasma density, the SFI rate and the strong-field-induced conductivity into a cycle-averaged and a sub-cycle component is depicted in Fig.~\ref{fig:decomposition}. While the cycle-averaged contributions have only positive values, the sub-cycle components can be both positive or negative. \\
As can be seen from Eq.~\ref{eq:decom_rho}, all harmonic orders of the two-color wave-mixing signal exhibit the same temporal signature given by $B(t)$. This is supported by the identical temporal profile of the detected harmonic peaks (see Fig.~\ref{fig:delay_projections}) and allows the extraction of the cycle-averaged, strong-field-induced plasma dynamics from an arbitrary order of the injection harmonics.  

\subsection{Test of the reconstruction procedure with the help of numerical simulations}
\begin{figure*}[htbp]
\centering
\includegraphics[width=\linewidth]{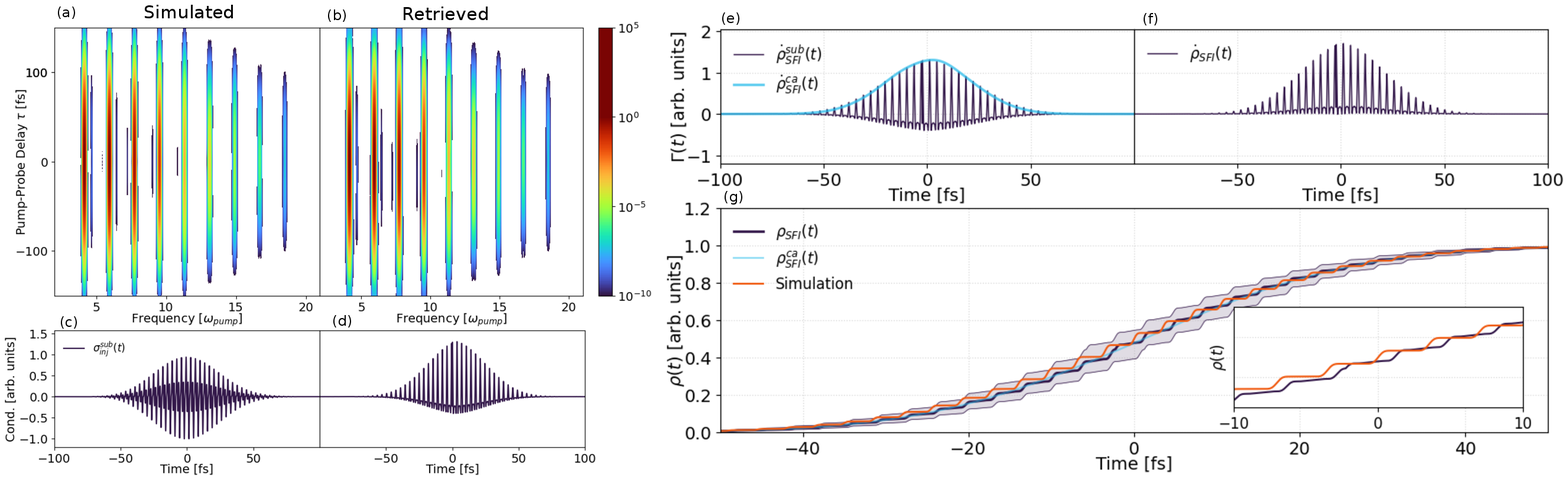}
\caption{Numerical test of the reconstruction procedure. (a). Simulated time-frequency map of the two-color wave-mixing signal generated by the injection current. (b). Retrieved spectrogram by the ptychographic iterative engine after 1000 iterations. (c) Direct output of the phase retrieval algorithm for $\sigma_{\rm{inj}}(t)$. (d). Corrected effective strong-field-induced conductivity after levelling the spectral phase. (e). Cycle-averaged and sub-cycle ionization rate computed from the corrected $\sigma_{\rm{inj}}(t)$. (f). Total SFI rate $\dot{\rho}_{\rm{SFI}}(t)$ obtained from a ptychographic analysis of simulated, two-color injection harmonics. (g). Resulting plasma dynamics induced by SFI compared to the numerical input data.}
\label{fig:simu_retrieval}
\end{figure*}

Figure~\ref{fig:simu_retrieval} shows a numerical test of the reconstruction procedure using numerical simulations of the optical signature of the injection current according to Eq.~\ref{eq:measured_quantity}. In the numerical model an instantaneous Ammosov-Delone-Krainov (ADK) tunneling rate \cite{Ammosov1986} is employed to calculate $\dot{\rho}_{\rm{SFI}}(t)$. The resulting time-resolved spectrum $Y_{\rm{inj}}(\omega, \tau)$ containing the first eight wave-mixing orders is displayed in Fig.~\ref{fig:simu_retrieval}(a). Figure~\ref{fig:simu_retrieval}(b) shows the reconstructed spectrogram of the ptychographic iterative engine after 1000 iterations. The spectral features of the input data are accurately reproduced and lead to the retrieved effective conductivity $\sigma_{\rm{inj}}(t)$ shown in Fig.~\ref{fig:simu_retrieval}(c). Analogous to the case of the experimental harmonic spectra the harmonic orders do not spectrally overlap and therefore the phase retrieval algorithm is not able to determine the relative phase of the harmonics. In the case of simulated data where the absolute and relative phases of all fields are well-known and the SFI process is assumed to be instantaneous we can manually set the spectral phase over the whole frequency range to 0. This leads to the sub-cycle modulated effective conductivity displayed in Fig.~\ref{fig:simu_retrieval}(d). \\
The resulting $\sigma_{\rm{inj}}^{\rm{sub}}(t)$ is used to reconstruct the sub-cycle and the cycle-averaged component of the SFI rate shown in Fig.~\ref{fig:simu_retrieval}(e) while the corresponding total SFI rate is depicted in Fig.~\ref{fig:simu_retrieval}(f). Direct integration of the obtained $\dot{\rho}_{\rm{SFI}}(t)$ leads to the expected stepwise increase of the plasma density [$\rho_{\rm{SFI}}(t)$ in Fig.~\ref{fig:simu_retrieval}(g)] that is compared to the cycle-averaged plasma formation and the numerical input of the simulation. The numerical input of the simulation is reproduced with a very high precision demonstrating that the reconstruction of $\rho_{\rm{SFI}}(t)$ from a time-resolved measurement of low-order injection harmonics is indeed feasible. Note that the shift between the simulated and reconstructed curve results from the CEP-insensitivity of the retrieval algorithm that is accounted for by the uncertainty of one optical cycle indicated by the shaded area in Fig.~\ref{fig:simu_retrieval}(g). Hence, the presented experimental strategy enables the reconstruction of sub-cycle ionization dynamics with attosecond precision if the relative phase of the harmonic orders is known.
\begin{figure}[htbp]
\centering
\includegraphics[width=\linewidth]{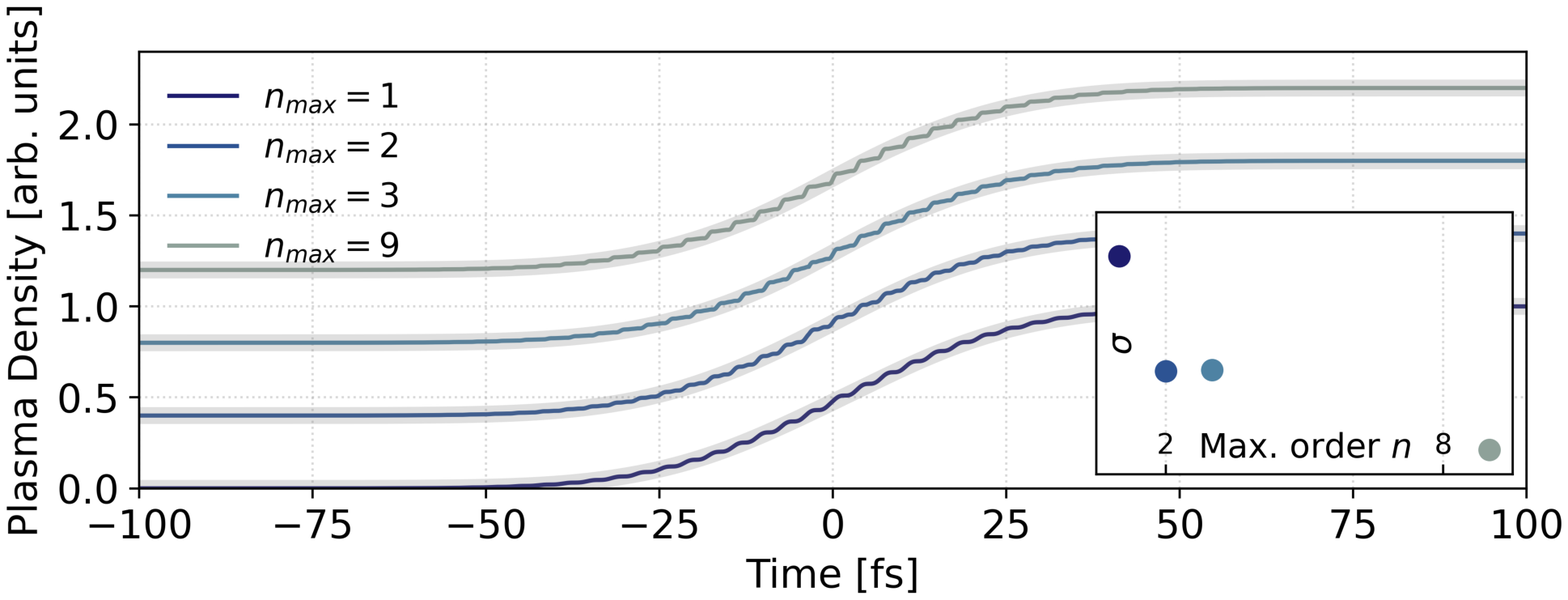}
\caption{Influence of the maximum harmonic order $n$ on the reconstruction of $\rho_{\rm{SFI}}(t)$ from simulated, time-resolved spectra (results for $n\geq 2$ are vertically shifted for visibility). Inset: Standard deviation between the individual reconstructed densities and the numerical input data.}
\label{fig:influence_of_n}
\end{figure}
Analysis of the influence of the maximum harmonic order $n_{\rm{max}}$ on the reconstruction of the SFI-induced plasma density is analyzed in Fig.~\ref{fig:influence_of_n}. Herein, the plasma densities reconstructed from more than the first harmonic order are shifted vertically for better visibility. The inset shows the standard deviation of the individual reconstructed plasma densities with respect to the numerical input data. These results indicate that the reconstruction of the sub-cycle dynamics of SFI benefits from an increased number of harmonics inserted into the retrieval algorithm. However, orders with $n>1$ only lead to minor modifications of the reconstructed ionization dynamics. Hence, analyzing the $n=1$ harmonic only already gives valuable insights into the strong-field-driven plasma dynamics while the exact shape of the nearly-stepwise increase of $\rho_{\rm{SFI}}(t)$ is encoded in the higher harmonic orders. 
\subsection{Case of dominating Brunel harmonics}
If in contrast to the above discussion the Brunel current instead of the injection current becomes the dominant source of low-order harmonic generation, it is still possible to reconstruct the strong-field-induced plasma formation from a time-resolved measurement of the two-color harmonics. The expression for the current density given in Eq.~\ref{eq:injection_current} needs to be replaced by the Brunel current $j_{\rm{Br}}(t)$ \cite{Brunel1990, Babushkin2017}:
\begin{equation}
\label{eq:Brunel_current}
\frac{\partial j_{\rm{Br}}(t, \tau)}{\partial t} = \frac{q_{e}^{2}}{m_e} \rho_{\rm{SFI}}(t) E_{\rm{probe}}(t-\tau) 
\end{equation}
Again, the above equation is only valid for cross-polarized pump and probe laser fields where the ionization is solely driven by the strong pump laser field. In analogy to the derivation of Eq.~\ref{eq:measured_quantity} the resulting spectrogram can be written as
\begin{align}
Y_{\rm{Br}}(\omega, \tau) &= \left \vert \int \frac{q_{e}^2}{m_e} \rho_{\rm{SFI}}^{\rm{sub}}(t) E_{\rm{probe}}(t-\tau) e^{i \omega t} dt \right \vert ^2 
\end{align}
Hence, phase-retrieval analysis of $Y_{\rm{Br}}(\omega, \tau)$ gives direct access to the strong-field-induced, sub-cycle variations of the plasma density $\rho_{\rm{SFI}}^{\rm{sub}}(t)$. The total plasma density due to SFI can be obtained in an analogous fashion to the previously described case of dominating injection harmonics.

\end{document}